\documentclass[preprintnumbers,showpacs,amsmath,amssymb,floatfix,9pt,prd,twocolumn,
superscriptaddress,nofootinbib]{revtex4}
\usepackage{CJK}
\usepackage{graphicx}
\usepackage{latexsym}
\usepackage{epsfig}
\usepackage{enumerate}
\usepackage{amssymb}
\usepackage[utf8]{inputenc}
\begin{document}
\title{(3+1)-D gravastar in de Rham–Gabadadze–Tolley like massive
gravity }

\author{Piyali Bhar}
\email{piyalibhar90@gmail.com}
\affiliation{Department of
Mathematics,Government General Degree College Singur, Hooghly, West Bengal 712 409,
India}

\begin{abstract}
The physical properties of a (3+1)-D gravastar in the context of massive gravity are discussed in this work.  In present investigation, the field equations have been solved for a static, uncharged sphere in order to achieve the gravastar model as proposed by Mazur and Mottola [Mazur and Mottola in Report No. LA-UR-01-5067,(2001); Mazur and Mottola, {\em Proc Natl Acad Sci} USA 101:9545, (2004)]. We address length of thin shell, energy, and entropy for the thin shell containing an ultra-relativistic stiff fluid. Israel matching criteria are used to ensure that the inner and outside geometries join smoothly. It turns out that the behavior of the gravastar is entirely altered by the existence of the graviton mass. Particularly, when $m\rightarrow0$, our findings precisely matched the outcomes of general relativity.
\end{abstract}

\maketitle

\section{Introduction}
Mazur and Mottola \cite{Mazur:2004fk,Mazur:2001fv} hypothesized the model of gravastar as an alternate structure to a black hole. The internal and external domains along with the shell region acting as an intermediary between them constitute the three sectors of the gravastar geometry. For each of these three zones, three different forms of the equation of state parameters are defined. According to Horvat and Ilijic \cite{Horvat:2007qa} the gravastar model contains a single spherical delta-shell at which the vacuum phase transition occurs between the de Sitter interior and the Schwarzschild or Schwarzschild-de Sitter outer geometries. By assuming that the dominant energy condition (DEC) will hold at the shell, they are able to estimate sharp analytic bounds on the surface compactness $2m/r$. The previous work was generalized by Horvat et al. \cite{Horvat:2008ch} by introducing an electrically charged component. The Einstein-Maxwell field equations are solved in the asymptotically de Sitter interior, where the fluid energy density is connected to the source of the electric field. Two distinct levels of complexity are applied to the linked nonlinear issue in order to demonstrate the possibility of finding solutions that possess all the characteristics of a gravastar in circumstances when the prerequisites for their existence are being pushed to their limit. In an another paper, they used the traditional Chandrasekhar method to study the radial stability of the continuous pressure gravastar \cite{Horvat:2011ar}. The authors solve Einstein equations for small perturbations around the equilibrium as an eigenvalue problem for radial pulsations and extract the equation of state for the static gravastar solutions. Authors have demonstrated that a splitting point between stable and unstable gravastar configurations is represented by the extremum of the $\rho_c-R$ curve for the centre energy density. The work of Chan et al. \cite{Chan:2009hp} demonstrated that depending on the total mass of the system, cosmological constants, the equation of state of the thin shell, and the initial position of the dynamical shell, the final output can be a black hole, stable gravastar, ``bounded excursion" stable gravastar, or de Sitter spacetime. The authors have discovered that the gravastar formation is constrained by the exterior cosmological constant; that is, the exterior cosmological constant needs to be less than the interior cosmological constant. Furthermore, it has been demonstrated that a black hole can still arise in the specific scenario where the Schwarzschild mass disappears notwithstanding the inability to build a stable gravastar. Recently, Ghosh et al. \cite{Ghosh:2023wps} presented a typical gravastar model within the context of UV corrected Loop Quantum Cosmology. According to the authors' findings, a bounce mechanism can prevent the centre singularity of a self-gravitating object, causing the gravastar's inner density to achieve a maximum critical density and then remain there because of an acting repulsive force. In agreement with the conjecture made by Mazur and Mottola, Nojiri and Nashed \cite{Nojiri:2023zlp} provide a stellar model with two scalar fields within the context of Einstein's gravity. To prevent ghosts, the two scalar fields in the model impose constraints that cause them to become non-dynamical. The authors of this model postulated that the Schwarzschild spacetime makes up the outside region and the de Sitter spacetime makes up the inside. They make the assumption that a polynomial function of the radial coordinate provides the metric on the shell that connects the two sections to one another. The concept of a gravitational Bose-Einstein condensate star in the presence of charge has been studied by Bhattacharjee and Chattopadhyay \cite{Bhattacharjee:2023apw} who have investigated the function of charge in the formation of gravastars and their characteristics in cylindrically symmetric space-time. With this method, thin shell approximation takes the place of the event horizon and eliminates the possibility of singularity at the gravastar's centre. Within the context of energy-momentum squared gravity, Sharif and Naz \cite{Sharif:2022ibt} examined the impact of charge on the physical characteristics of gravastars. Specifically, they found that the charge within the inner core of gravastars maintains equilibrium by counteracting the inward gravitational force. The impact of the Kuchowicz metric function on the gravastar model within the context of braneworld gravity was explained by Ray et al. \cite{Ray:2022lxl}.  As an alternative for the General Relativity (GR), Sengupta et al. \cite{Sengupta:2020lhw} and Arba\~nil et al. \cite{Arbanil:2019xfi} derived stable gravastar models within the framework of the Randall-Sundrum (RS) braneworld model which is particularly relevant in the context of cosmology and astrophysics. Within the context of the Mazur and Mottola model, Usmani et al. \cite{Usmani:2010ac} have developed a charged (3 + 1)-dimensional gravastar that admits conformal motion. The authors offer a substitute for static black holes in this study. It is discovered that the energy density in the interior region of the gravastar diverges. Regretfully, this leads to the model being singular at r = 0, Bhar's work \cite{Bhar:2014vra} provides an extension of Usmani et al. \cite{Usmani:2010ac} on charged gravastar, admitting conformal motion with higher dimensional space-time. The four-dimensional work on gravastar by Usmani et al. \cite{Usmani:2010ac} was generalized to higher-dimensional space-time by Ghosh et al. \cite{Ghosh:2015ohi}, although conformal motion was not permitted in their studies. The authors' primary goal was to develop gravastars in Einstein-Maxwell geometry and observe the higher-dimensional impacts. They also demonstrated that there is a significant amount of quantitative change in the physical parameter profiles using graphical analysis. As one progresses towards higher dimensions, notable outcomes become apparent. Further papers on gravastar in Einstein gravity as well as in modified gravity are available in \cite{Errehymy:2017wpc,Pani:2015tga,Chirenti:2016hzd,Rutkowski:2020fvm,Das:2020lqy,Yousaf:2020nza,Das:2017rhi,Barzegar:2023ueo,Rahaman:2012wc,Ghosh:2019nsi, Ghosh:2019upa,Rahaman:2012xx,DeBenedictis:2005vp}.\\

Fierz and Pauli \cite{Fierz:1939ix} successfully generated massive gravity for the first time by taking non-zero mass gravitons into account in 1939. Subsequent research such as van Dam-Veltman-Zakharov discontinuity\cite{VanNieuwenhuizen:1973fi, vanDam:1970vg,Zakharov:1970cc}, Vainshtein mechanism \cite{Vainshtein:1972sx}  and Boulware-Deser ghost \cite{Boulware:1972yco}, etc, however, revealed that this modified gravity theory contains several difficulties. The de-Rham-Gabadadze-Tolley (dRGT) massive gravity theory was presented in 2010 \cite{deRham:2010ik, deRham:2010kj}, and it effectively resolved the issues mentioned above. Since the dRGT massive gravity gives the graviton mass, it is a logical extension of Einstein's theory of general relativity and a fantastic field for theoretical physics study. The dRGT massive gravity admits general relativity as a special case, explaining nonlinear interaction terms as a correction of the Einstein-Hilbert action. It is hypothesized that dRGT massive gravity may provide a plausible explanation for the accelerated expansion of the Universe that doesn't need any dark energy or cosmological constant. Without taking into account any extra dark matter, the dRGT model is quite satisfactorily fitted with observational data for LSB galaxies and Milky Way rotation curves in \cite{Panpanich:2018cxo}. In this context, it has also been observed that the dRGT model is consistent with the NFW profile. In dRGT massive gravity, the spherically symmetric solutions were covered in \cite{Nieuwenhuizen:2011sq, Brito:2013xaa},The associated charged black hole solution was discovered in \cite{Berezhiani:2011mt}, and its bi-gravity extension was discovered in \cite{Babichev:2014fka}, including as specific cases the previously identified spherically symmetric black hole solutions.
The massive gravity theory has acquired substantial interest in the theoretical realm due to the detection of gravitational waves caused by the merging of two Black holes and massive stars, as detected by LIGO and VIRGO \cite{LIGOScientific:2016aoc, LIGOScientific:2017vwq}.

In this work, we shall examine gravastar solutions in dRGT like massive gravity, and investigate different physical properties. The structure of the paper is as follows: We define the action and underlying field equations in dRGT like massive gravity in Sec \ref{sec2}. In Sec \ref{sec3}, we have obtained the model of the gravastar by solving the field equations in three different regions of the gravastar. The purpose of Sec \ref{sec4} is to explain the junction condition. The impact of the mass of the graviton on the physical quantities of the gravastar model is discussed via graphical analysis in the next section. Sec \ref{sec6} describes the stability of the gravastar. Contour plots are used to illustrate some physical characteristics of the gravastar model in Sec \ref{sec7}, and the last section provides some closing remarks.

\section{action and field equation in massive gravity}\label{sec2}
In dRGT like massive gravity, the action $\mathcal{I}$ is defined by,
\begin{equation}
\mathcal{I}=\frac{1}{16\pi G}\int d^{4}x\sqrt{-g}\left[ \mathcal{R}
+m^{2}\sum_{i}^{4}c_{i}\mathcal{U}_{i}(g,f)\right] +I_{\text{matter}},
\label{Action}
\end{equation}%
where $I_{\text{matter}}$ is the matter Lagrangian, $g$ and $f$ are a fixed symmetric tensor and a metric tensor, respectively, and $\mathcal{R}$ is the Ricci scalar curvature. The parameter $m$ is associated with the gravitational mass. $\mathcal{U}_i$ represents the effective potential for graviton,and it has the following expression in four dimensions of spacetime:
\begin{eqnarray}
\mathcal{U}_{1} &=&\left[ \mathcal{K}\right] ,\;\;\;\;\;\;\;\mathcal{U}_{2}=%
\left[ \mathcal{K}\right] ^{2}-\left[ \mathcal{K}^{2}\right] ,  \nonumber \\
\mathcal{U}_{3} &=&\left[ \mathcal{K}\right] ^{3}-3\left[ \mathcal{K}\right] %
\left[ \mathcal{K}^{2}\right] +2\left[ \mathcal{K}^{3}\right] ,  \nonumber \\
\mathcal{U}_{4} &=&\left[ \mathcal{K}\right] ^{4}-6\left[ \mathcal{K}^{2}%
\right] \left[ \mathcal{K}\right] ^{2}+8\left[ \mathcal{K}^{3}\right] \left[
\mathcal{K}\right] +3\left[ \mathcal{K}^{2}\right] ^{2}-6\left[ \mathcal{K}%
^{4}\right] .  \nonumber
\end{eqnarray}
where [$\mathcal{K}$] represents the trace of the metric matrix $\mathcal{K_{\nu }^{\mu }}=\sqrt{g^{\mu \alpha }f_{\alpha \nu }}$.\\
The equation of motion for this gravity can be found by varying the action (\ref{Action}) with respect to $g_{\nu}^{\mu}$ as follows:
\begin{equation}
R_{\mu \nu }-\frac{1}{2}R g_{\mu \nu }+m^{2}\chi _{\mu \nu}=\frac{8\pi G}{c^4}T_{\mu \nu },  \label{Field equation}
\end{equation}%
where the expression for $\chi _{\mu \nu } $ is given by,
\begin{eqnarray}
\chi _{\mu \nu } &=&-\frac{c_{1}}{2}\left( \mathcal{U}_{1}g_{\mu \nu }-%
\mathcal{K}_{\mu \nu }\right)\nonumber\\&& -\frac{c_{2}}{2}\left( \mathcal{U}_{2}g_{\mu
\nu }-2\mathcal{U}_{1}\mathcal{K}_{\mu \nu }+2\mathcal{K}_{\mu \nu
}^{2}\right)  \nonumber \\
&&  \nonumber \\
&&-\frac{c_{3}}{2}(\mathcal{U}_{3}g_{\mu \nu }-3\mathcal{U}_{2}\mathcal{K}%
_{\mu \nu }+6\mathcal{U}_{1}\mathcal{K}_{\mu \nu }^{2}-6\mathcal{K}_{\mu \nu
}^{3})  \nonumber \\
&&  \nonumber \\
&&-\frac{c_{4}}{2}\left( \mathcal{U}_{4}g_{\mu \nu }-4\mathcal{U}_{3}%
\mathcal{K}_{\mu \nu }+12\mathcal{U}_{2}\mathcal{K}_{\mu \nu }^{2}\right.\nonumber\\&&\left.-24
\mathcal{U}_{1}\mathcal{K}_{\mu \nu }^{3}+24\mathcal{K}_{\mu \nu
}^{4}\right).
\end{eqnarray}
Now to describe a gravastar spacetime in $(3+1)$-dimension, let us consider the following line element,
\begin{equation}\label{line}
ds^{2}=-e^{2\phi(r)}dt^{2}+e^{2\lambda(r)}dr^{2}+r^{2}(d\theta^2+\sin^2\theta d\phi^2),
\end{equation}
where $\phi$ and $\lambda$ are purely radial, i.e., they only depend on `r'. We use the following reference metric, generalizing and adhering to the ansatz in \cite{Cai:2014znn}:
\begin{equation}
f_{\mu \nu }=\text{diag}(0,0,C^{2},C^{2}\sin ^{2} \theta ),  \label{f11}
\end{equation}
where $C$ is a positive constant. With the reference metric (\ref{f11}), we have \cite{Cai:2014znn},
\begin{equation}
\mathcal{U}_{1}=\frac{2C}{r},\text{ \ \
}\mathcal{U}_{2}=\frac{2C^{2}}{r^{2}},\,\mathcal{U}_{3}=0,\,\mathcal{U}_{4}=0. \nonumber
\end{equation}
Here, we consider the gravastar model having perfect fluid distribution whose energy-momentum tensor is given by,
\begin{equation}
T_{\mu \nu }=\left( c^{2}\rho +p\right) U_{\mu }U_{\nu }-pg_{\mu \nu },
\label{EMTensorEN}
\end{equation}
where $U_{\mu }$ is the fluid four-velocity. $p$ and $\rho $ are the fluid pressure and density, respectively, as measured by the local observer.
Furthermore, considering equations (\ref{EMTensorEN}) and (\ref{line}), it is also straightforward to obtain the subsequent nonzero field equation components:
\begin{eqnarray}
\frac{8\pi G}{c^{2}}r^{2}\rho &=&e^{-2\lambda}(2r\lambda'-1)+1+m^2C(c_2C+c_1r),  \label{1} \\
&&  \nonumber \\
\frac{8\pi G}{c^{4}}r^{2}p &=&e^{-2\lambda}(2r\phi'+1)-1-m^2C(c_2C+c_1r),  \label{2} \\
&&  \nonumber \\
\frac{8\pi G}{c^{4}}r p &=&e^{-2\lambda}(\phi'+r\phi'^2-\lambda'-r\lambda'\phi'+r\phi'')-\frac{m^2c_1C}{2}.  \label{3}\nonumber\\
\end{eqnarray}%
Here the the first and
the second derivatives of a function with respect to $r$ respectively denoted by prime and double prime.\\
The conservation equation can be obtained from $T_{\mu;\nu}^{\nu}$=0 as follows:
\begin{equation}
\frac{dp}{dr}+\left( c^{2}\rho +p\right) \frac{d\phi}{dr}=0.
\label{extraEQ}
\end{equation}
To find the expression of $e^{-2\lambda}$, we can use Eq. (\ref{1}) which leads to
\begin{equation}
e^{-2\lambda}=1-\frac{2G}{c^2r}\int4\pi r^2\rho dr+m^{2}C\left( \frac{c_{1}r}{2}+c_{2}C\right).  \label{4g(r)}
\end{equation}%
From eqn.(\ref{2}), the expression for $\phi'$ can be obtained as,
\begin{eqnarray}\label{tov}
\frac{d\phi}{dr}&=&\left[\frac{4\pi G}{c^{4}}r p +\frac{1+m^2C(c_2C+c_1r)}{2r}-e^{-2\lambda}\right]e^{2\lambda}.\nonumber\\
\end{eqnarray}
We will solve the field equations in the upcoming section to obtain the gravastar model.


\section{The geometry of a Gravastar}\label{sec3}
In case of gravastar, there are three distinct zones, and each has a specific width. These zones are represented by the geometrical regions $b = r_1<r<r_2 = b + \epsilon$, where $\epsilon$ is a very small positive quantity and $r_1$ and $r_2$ stand for the radii of the interior and exterior regions, respectively.
\begin{itemize}
  \item Interior de Sitter region: $\rho=-p$, $r\in [0,\,r_1)$
  \item Thin Shell: $\rho=p$, $r\in (r_1,\,r_2)$
  \item Exterior Schwarzschild geometry: $\rho=p=0$, $r\in (r_2,\,\infty)$
\end{itemize}
The metric functions $e^{2\phi}$ and $e^{2\lambda}$ must be continuous at the interfaces $r = r_1$ and $r = r_2$, but the first derivatives of $e^{2\phi}$ and $e^{2\lambda}$, and $p$ must be discontinuous \cite{Mazur:2001fv}.
\subsection{Interior region}
Following Mazur-Mottola \cite{Mazur:2001fv}, we now take into consideration the following equation of state (EoS)
\begin{equation}\label{f4}
p=-\rho,
\end{equation}
in order to solve the above field eqns. (\ref{1})-(\ref{3}) within the interior region of the gravastar. It is the generalized version of the dark energy EoS with $\omega = -1$. The inward gravitational attraction of the shell is balanced by this negative pressure, which operates radially outward from the center of the spherically symmetric gravitating system.\\
Solving eqn.(\ref{tov}) with the help of (\ref{f4}), we obtain,
\begin{eqnarray}\label{e7}
\rho=-p=\rho_c,
\end{eqnarray}
where $\rho_c$ is a constant and this constant is denoted by $\frac{3H^2}{8\pi G_N}$ \cite{Mazur:2001fv}.\\
By choosing $G=c=1$ and $8\pi=k$ (from here and onwards), from eqns. (\ref{1}) and (\ref{e7}) we get the following equation:
\begin{equation}\label{eq1}
kr^2 \rho_c=e^{-2\lambda}(2r \lambda'-1)+1+m^2C(c_2C+c_1r).
\end{equation}
On integrating the above equation we obtain the expression for $e^{-2\lambda}$ as,
\begin{eqnarray}
e^{-2\lambda}=1+m^2C(c_2C+\frac{c_1r}{2})-k\rho_c\frac{r^2}{3}+\frac{D}{r},
\end{eqnarray}
where $D$ is the constant of integration. One can require that $D=0$ in order to make the solution regular at the origin, which provides the expression of the metric coefficients as,
\begin{eqnarray}\label{k1}
e^{-2\lambda}=1+m^2C(c_2C+\frac{c_1r}{2})-k\rho_c\frac{r^2}{3}.
\end{eqnarray}
Adding eqns. (\ref{1}) and (\ref{2}) we get,
\begin{eqnarray}\label{e20}
k(p+\rho)r^2=e^{-2\lambda}2r(\lambda'+\phi'),
\end{eqnarray}
Solving (\ref{e20}) with the help of (\ref{f4}) we obtain,
\begin{eqnarray}\label{s1}
\lambda'+\phi'=0.
\end{eqnarray}
Integrating the eqn.(\ref{s1}), the expression for $e^{2\phi}$ can be obtained as follows:
\begin{eqnarray}\label{k2}
e^{2\phi}&=&A e^{-2\lambda},\nonumber\\
\Rightarrow~e^{2\phi}&=&A\Big[1+m^2C(c_2C+\frac{c_1r}{2})-k\rho_c\frac{r^2}{3}\Big],
\end{eqnarray}
where $A$ is constant of integration. Here, one can note that both the metric potentials depend on the massive gravity parameters $m^2c_1$ and $m^2c_2$.\par
One can avoid the issue of the core singularity of a classical black hole by observing that the interior solutions of the aforementioned solutions (\ref{k1}) \& (\ref{k2}) do not have a singularity.\\
One can derive the active gravitational mass $M(r)$ using the following formula:
\begin{eqnarray}
M(r)&=&4\pi\int_0^{r}\eta^{2}\rho(\eta)d\eta\nonumber\\
&=&\frac{4\pi r^3}{3}\rho_c.
\end{eqnarray}

\subsection{The Intermediate Thin Shell}
According to the hypothesis proposed by Mazur \& Mottola \cite{Mazur:2001fv}, the relationship between the pressure and the energy density in the shell of the gravastar is,
\begin{equation}\label{e2}
p=\rho.
\end{equation}
Getting a general solution to the field equations in the non-vacuum region, or inside the shell, is challenging. We search for an analytical solution between $0 < (e^{-2\lambda}\equiv h) << 1$ in the thin shell limit. We can set $h$ to zero to the leading order as a benefit of it.Using this approximation, the field equations (8)-(10) with the aforementioned EoS can be reformulated in the following way:
\begin{eqnarray}
kr^2\rho&=&e^{-2\lambda}(2r\lambda')+1+m^2C(c_2C+c_1r),\label{e3}\\
kr^2p&=&-1-m^2C(c_2C+c_1r),\label{e4}\\
krp&=&-e^{-2\lambda}\lambda'-r\lambda'\phi'e^{-2\lambda}-\frac{m^2c_1C}{2}\label{e5}
\end{eqnarray}
Solving the eqns. (\ref{e3})-(\ref{e4}) with the help of eqn. (\ref{e2}), we obtain the expression for $e^{-2\lambda}$ as,
\begin{eqnarray}
e^{-2\lambda}=2\ln r+2m^2C(c_2C\ln r+c_1r)+B,
\end{eqnarray}
From eqns. (\ref{e4}) and (\ref{e5}), with the help of the expression $e^{-2\lambda}$, we obtain the expression of the another metric coefficient as,
\begin{eqnarray}
  e^{2\phi} &=& E\left(\frac{1+m^2C^2c_2+m^2Cc_1r}{r^4}\right).
\end{eqnarray}
From the TOV equation, with the help of (\ref{e2}), we get,
\begin{eqnarray}\label{e6}
  \frac{dp}{dr} &=& -2p\frac{d\phi}{dr},
\end{eqnarray}
Solving (\ref{e6}), we obtain the expression of pressure inside the thin shell as,
\begin{eqnarray}
p&=&Fe^{-2\phi},\\
&=&\frac{Gr^4}{1+m^2C^2c_2+m^2Cc_1r}=\rho,
\end{eqnarray}
where, both $F$ and $G$ are constants of integration.\\
\begin{figure}[htbp]
        \includegraphics[scale=.6]{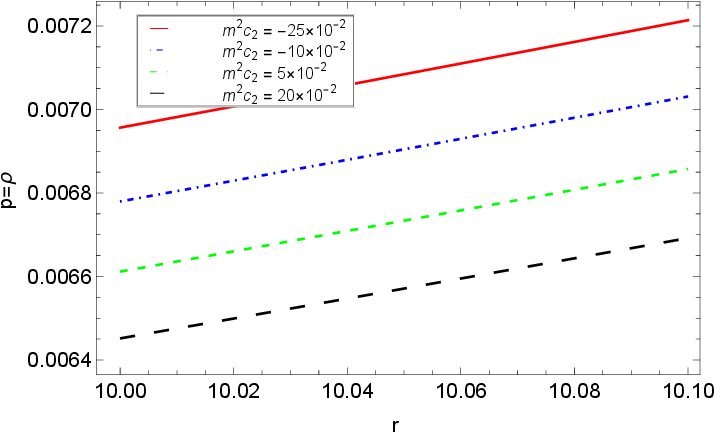}
         \includegraphics[scale=.6]{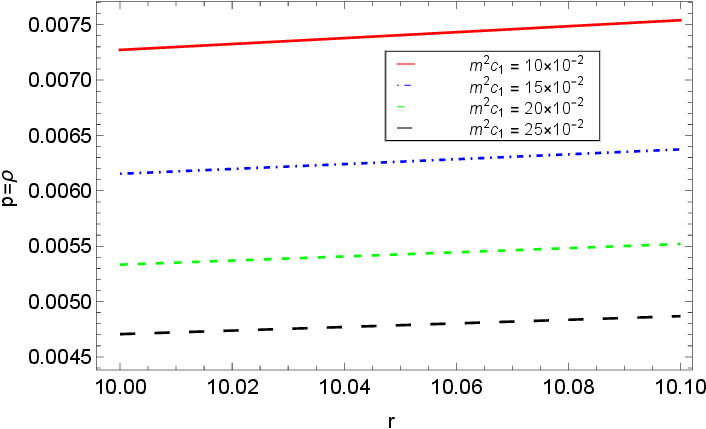}
       \caption{Pressure and density variations within the thin shell
\label{sp}}
\end{figure}

\subsection{Exterior Spacetime and matching}
We take into consideration $ p=\rho=0 $ for this region, which guarantees that the Schwarzschild line element describes the outer space-time which is described by,
\begin{eqnarray}
ds^{2}&=& \left(1-\frac{2\mathcal{M}}{r}\right) dt^{2}-\left(1-\frac{2\mathcal{M}}{r}\right)^{-1}dr^{2}\nonumber\\&&-r^{2}(d\theta^{2}+\sin^{2}\theta d\phi^{2}),\label{exterior}
\end{eqnarray}
where, $\mathcal{M}$ being the mass of the gravastar.
The gravastar structure, which resembles a hollow sphere, features two junction surfaces as opposed to the single junction surface found in a compact star. One is at $r = r_1$, between the internal area and the intermediate thin shell, while the other is at $r = r_2$, between the shell and external spacetime. By using the continuity of the metric potentials $g_{rr}$ and $g_{tt}$ for our current gravastar model at the junction of the core and shell at $r = r_1$ the following relationships are obtained.
\begin{widetext}
\begin{eqnarray}\label{m1}
1+m^2C(c_2C+\frac{c_1r_1}{2})-k\rho_c\frac{r_1^2}{3}&=&2\ln r_1+2m^2C(c_2C\ln r_1+c_1r_1)+B,\\
A\left[1+m^2C(c_2C+\frac{c_1r_1}{2})-k\rho_c\frac{r_1^2}{3}\right]&=& E\left(\frac{1+m^2C^2c_2+m^2Cc_1r_1}{r_1^4}\right)
\end{eqnarray}
\end{widetext}
Now using the matching condition between the shell and the
exterior region at $r = r_2$ (exterior radius) yields the following relationships:
\begin{eqnarray}
2\ln r_2+2m^2C(c_2C\ln r_2+c_1r_2)+B&=&1-\frac{2\mathcal{M}}{r_2},\label{m2}\nonumber\\
\\
E\left(\frac{1+m^2C^2c_2+m^2Cc_1r_2}{r_2^4}\right)&=&1-\frac{2\mathcal{M}}{r_2}.\label{m3}\nonumber\\
\end{eqnarray}
In the current work, in order to examine the physical behavior of various model parameters for the gravastar model, we have to choose values for the unknown parameters. To determine the numerical values of three different constants $B,\,E,\, A$, and the mass of the gravastar, we choose interior radius $r_1= 10$ km., while the outer radius is $10.1$ km. i.e., the thickness of the thin shell is equal to $0.1$ km. \\
Using a specific choice of $r_1 = 10$ km., $r_2=10.1$ km, $m^2c_1=0.1$, $C=0.5$, $G = 0.1\times10^{-5}$, and $\rho_c = 0.001~ km^{-2}$, we simultaneously solved equations (\ref{m1})-(\ref{m3}) to determine the values of the constants $B,\,E,\, A$, and mass of the gravastar for four distinct values of $m^2c_2$. The results are shown in Table~\ref{table1}.\\
On the other hand, by simultaneously solving equations (\ref{m1})-(\ref{m3}) for a specific option of $r_1 = 10$ km., $r_2=10.1$ km, $m^2c_2=-0.5$, $C=0.5$, $G = 0.1\times10^{-5}$, and $\rho_c = 0.001~ km^{-2}$, we were able to calculate the values of the constants $B,\,E,\, A$, and mass of the gravastar for four distinct values of $m^2c_1$. The result has been presented in table~\ref{table2}.\\
From both the tables, one can note that the ratio of obtain mass of the gravastar ($\mathcal{M}$) and outer radius $r_2$  for all the combinations of $\{m^2c_1,\, m^2c_2\}$ satisfy the ratio $\frac{\mathcal{M}}{r_2}<\frac{4}{9}$, i.e., respect the buchdahl limit.


\begin{table}[t]
\centering
\caption{Values of $B,\,E,\,A$ and mass of the gravastar have been obtained for four distinct values of $m^2c_2$ with $r_1 = 10$ km., $r_2=10.1$ km, $m^2c_1=0.1$, $C=0.5$, $G = 0.1\times10^{-5}$, and $\rho_c = 0.001~ km^{-2}$}\label{table1}
\begin{tabular}{@{}ccccccccccccc@{}}
\hline
$m^2c_2$&&$B$&$M$&$E$& $A$\\
\hline
$-0.25$&& $-4.96761$&$3.13909$& $2729.73$ & $0.137238$\\
\hline
$-0.10$&&  $-5.1028$& $2.94594$ & $2929.48$ & $0.167326$\\
\hline
$0.05$&& $-5.23799$ & $2.7528$ & $3119.35$ & $0.200394$\\
\hline
$0.20$&& $-5.37319$ & $2.55965$ & $3300.07$ & $0.236442$\\
\hline
\end{tabular}
\end{table}

\begin{table}[t]
\centering
\caption{Values of $B,\,E,\,A$ and mass of the gravastar have been obtained for four distinct values of $m^2c_1$  with  $r_1 = 10$ km., $r_2=10.1$ km, $m^2c_2=-0.5$, $C=0.5$, $G = 0.1\times10^{-5}$, and $\rho_c = 0.001~ km^{-2}$}\label{table2}
\begin{tabular}{@{}ccccccccccccc@{}}
\hline
$m^2c_1$&&$B$&$M$&$E$& $A$\\
\hline
$0.10$&& $-4.74228$&$3.46099$& $2372.69$ & $0.0937112$\\
\hline
$0.15$&&  $-5.11728$& $2.80449$ & $2834.36$ & $0.189872$\\
\hline
$0.20$&& $-5.49228$ & $2.14799$ & $3172.35$ & $0.31956$\\
\hline
$0.25$&& $-5.86728$ & $1.49149$ & $3430.49$ & $0.48276$\\
\hline
\end{tabular}
\end{table}
\section{Junction condition}\label{sec4}
According to Mazur and Mottola, the shell of the gravastar is confined with the cold radiation fluid by the surface tensions at the timelike interfaces $r_1$ and $r_2$. These result from discontinuities in pressure and can be computed using the Lanczos-Israel junction conditions \cite{Israel:1966rt,Berezin:1987bc}. The expression for surface stress energy tensor is described as \cite{Israel:1966rt},
\begin{equation}
\mathcal{S}^{\xi}_{\zeta}=-\frac{1}{8\pi}(\kappa^{\xi}_{\zeta}-\delta^{\xi}_{\zeta}\kappa^{\varphi}_{\varphi}),
\end{equation}
where $\xi,\, \zeta$ take the value $t,\,\theta,\,\phi$. $\kappa_{\xi \zeta}$ has the expression
$\kappa_{\xi \zeta}=\left[K_{\xi \zeta}\right]^{+}-\left[K_{\xi \zeta}\right]^{-}$, and it represents the discontinuity in the extrinsic curvature $K_{\xi \zeta}$. The expressions for $K_{\xi \zeta}^{\pm}$ is represented as,
\begin{equation}
K_{\xi \zeta}^{\pm}=-n_{\nu}^{\pm}\left(\partial_{\zeta}e_{\xi}^{\nu}+\Gamma^{\nu}_{\alpha\beta}e_{\xi}^{\alpha}e_{\zeta}^{\beta}\right),
\end{equation}
where the unit normal vectors on the surface is described by $n_{\nu}^{\pm}$  with $n^{\nu}n_{\nu}=1$, and $e_{\xi}^{\alpha}=\frac{\partial x^{\alpha}}{\partial \chi^{\xi}}$, $ \chi^{\xi}$ describes the coordinate on the shell. $\Gamma^{\nu}_{\alpha\beta}$ represents the Christoffel symbols. The signs ``$-$" and ``$+$" denote the interior spacetime, and exterior spacetime, i.e., Schwarzschild spacetime, respectively.\par

The surface energy density ($\sigma$) and surface pressure ($\mathcal{P}$) for our present model of gravastar in massive gravity can be calculated as,
\begin{widetext}
\begin{eqnarray}
\sigma &=&-\frac{1}{4\pi b}\left[\sqrt{f} \right]_{-}^{+}\nonumber\\
&=&-\frac{1}{4\pi b}\Big[\sqrt{1-\frac{2\mathcal{M}}{b}}-\sqrt{1+m^2C(c_2 C+\frac{c_1 b}{2})-k\rho_c\frac{b^2}{3}}\Big],\label{sigma1}\nonumber\\
\\
\mathcal{P}&=&-\frac{\sigma}{2}+\frac{1}{16\pi}\left[\frac{f'}{\sqrt{f}}\right]_{-}^{+}\nonumber\\
&=&\frac{1}{8\pi b}\left[\frac{1-\frac{\mathcal{M}}{b}}{\sqrt{1-\frac{2\mathcal{M}}{b}}}-\frac{12 + 3 C m^2 (c_1 + 4 C c_2 + 2 c_1 b) - 4 k b (1 + b) \rho_c}{12\sqrt{1+m^2C(c_2 C+\frac{c_1 b}{2})-k\rho_c\frac{b^2}{3}}}\right]\label{p1}.
\end{eqnarray}
\end{widetext}

With the surface energy density equation, determining the mass of the surface of the thin shell ($m_{\text{s}}$) is now straightforward, which is given by,
\begin{eqnarray}\label{ms}
m_{\text{s}}&=&4\pi b^2\sigma,\nonumber\\
&=&b\bigg[\sqrt{1+m^2C(c_2 C+\frac{c_1 b}{2})-k\rho_c\frac{b^2}{3}}-\sqrt{1-\frac{2\mathcal{M}}{b}}\bigg].\nonumber\\
\end{eqnarray}
From the above expression of $m_{\text{s}}$, we get a suitable range of the radius of the gravastar as,
\begin{eqnarray}
2\mathcal{M}<b<\frac{3 c_1 C m^2 + \sqrt{
 9 c_1^2 C^2 m^4 + 48 k (1 + C^2 c_2 m^2) \rho_c}}{4k \rho_c}.\nonumber\\
\end{eqnarray}
It is interesting to note that, twice the mass of the gravastar acts as a lower limit of `b'.\\
Performing some mathematical computations, we determine the mass of the gravastar in terms of the thin shell mass as follows:
\begin{eqnarray}
\mathcal{M}&=&\frac{b}{2}\Big[-\frac{1}{2} C m^2 (2 C c_2 + c_1 b) + \frac{1}{3} k b^2 \rho_c-\left(\frac{m_s}{b}\right)^2 \nonumber\\&&+\frac{2m_s}{b}\sqrt{1+m^2C(c_2 C+\frac{c_1 b}{2})-k\rho_c\frac{b^2}{3}}\Big].
\end{eqnarray}

\section{Some physical properties of our present model}\label{sec5}
The purpose of this section is to investigate how the massive parameter affects the physical characteristics of a 4D gravastar in the context of massive gravity. We investigate here the proper length of the thin shell, the energy of the relativistic structure of 4D gravastars inside the shell, the entropy inside the shell, and the EoS parameter. Furthermore, we will use graphical representations to demonstrate our results.
\subsection{Proper length of the shell}
Mazur and Mottola postulated that the stiff fluid satisfying the EoS $p=\rho$ is located in between the lower and upper boundaries of the shell, where the boundaries are given at $r = b$ and $r = b+\epsilon$ respectively with $0~<\epsilon\ll 1$. One can use the following formula to find the proper length $ \mathcal{L}$ of the shell as,
\begin{eqnarray}\label{l0}
   \mathcal{L}&=& \int_{b}^{b+\epsilon}\frac{1}{\sqrt{e^{-2\lambda}}} dr,\nonumber\\
  &=& \int_{b}^{b+\epsilon}\frac{1}{\sqrt{ 2\ln r+2m^2C(c_2C\ln r+c_1r)+B}}.
\end{eqnarray}
To perform the above integral analytically is quite difficult due to its expression. Let us assume that there exists a function $g(r)$ such that, $\frac{d}{dr}g(r)=\frac{1}{\sqrt{e^{-2\lambda}}}$.\\
Using the above assumption,
\begin{eqnarray}\label{l0}
   \mathcal{L}&=& \int_{b}^{b+\epsilon}\frac{d}{dr}g(r) dr=g(b+\epsilon)-g(b),
\end{eqnarray}
Using the Taylor series expansion of $g(b+\epsilon)$ about the point `b' and retaining upto the second order term of $\epsilon$, (since $\epsilon<<1$),the expression of $ \mathcal{L}$ can be obtained as,
\begin{eqnarray}\label{l0}
   \mathcal{L}&=& \frac{\epsilon}{\sqrt{ 2\ln b+2m^2C(c_2C\ln b+c_1b)+B}}\nonumber\\&&-\frac{\epsilon^2}{2}\frac{C m^2 (c_1 + \frac{C c_2}{b}) + \frac{1}{b}}{\left\{B + 2 \log(b) + 2 C m^2 \big(c_1 b + C c_2 \log(b)\big)\right\}^{\frac{3}{2}}}.\nonumber\\
\end{eqnarray}
\begin{figure}[htbp]
        \includegraphics[scale=.6]{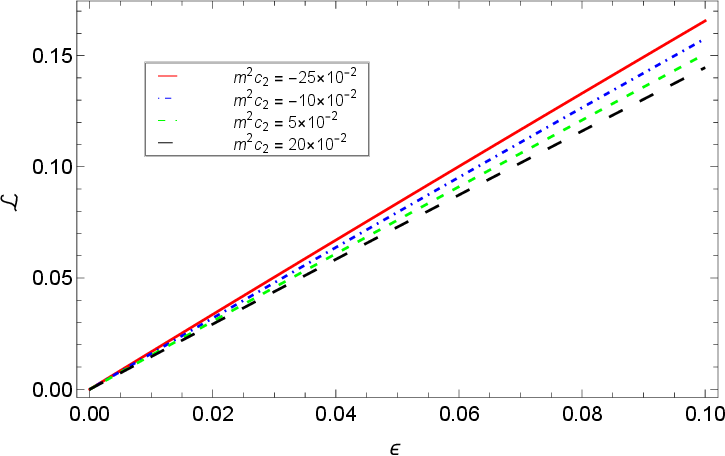}
        \includegraphics[scale=.6]{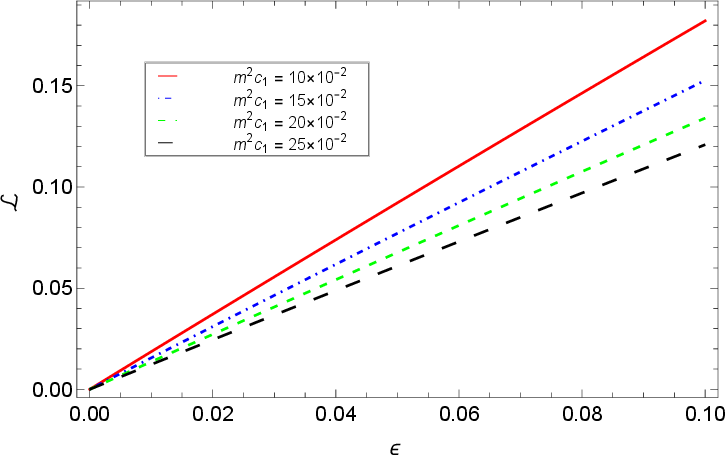}
       \caption{Variation between the gravastar's proper length inside its shell and its shell thickness.
\label{ell}}
\end{figure}
By retaining a fixed value for $m^2c_2$, the behavior of the proper thickness of the shell with respect to its thickness $\epsilon$ is depicted in Fig.~\ref{ell} for different values of $m^2c_1$ and vice versa. From the figures it is clear that proper length increases with thickness but for higher values of massive gravity parameters $m^2c_2$ and $m^2c_1$, it takes lower values.

\subsection{Entropy}
The measurement of entropy is correlated with the degree of disorder in a geometric structure. Thin-shell gravastar entropy is examined to comprehend the chaotic nature of gravastar geometry. Using the equation of state, $p=\rho=\frac{\alpha^2}{8\pi}\left(\frac{K_B}{\hbar}\right)^2$, one may get the entropy of the thin shell \cite{Mazur:2001fv}. According to the conventional thermodynamic Gibbs relation, $Ts = p + \rho$, for a relativistic fluid with zero chemical potential, the local specific entropy density is $$s(r)=\frac{2\alpha^2 K_B^2 T(r)}{\kappa\hbar^2}=2\alpha \left(\frac{K_B}{\hbar}\right)\sqrt{\frac{p(r)}{8\pi}}$$ at the temperature $T(r)$. In the above equation, $\alpha$ is a dimensionless constant, $K_B$ denotes the Boltzmann constant, and $h$ is the Planck constant and $\hbar=\frac{h}{2\pi}$.\\
We determine the entropy of the model as follows:
\begin{eqnarray}\label{l5}
\mathcal{S}&=&\int_{b}^{b+\epsilon}4\pi r^2 s(r)\sqrt{e^{2\lambda}}dr,\nonumber\\
&=&\sqrt{8\pi}\alpha \left(\frac{K_B}{\hbar}\right)\int_b^{b+\epsilon}r^2\sqrt{\frac{p(r)}{e^{-2\lambda}}} dr,
  \end{eqnarray}
By employing the expressions for $p$ and $e^{2\lambda}$ in eqn. (\ref{l5}) we get,
\begin{widetext}
\begin{eqnarray}\label{l6}
\mathcal{S}&=&\sqrt{8\pi}\alpha \left(\frac{K_B}{\hbar}\right)\int_b^{b+\epsilon}r^2\sqrt{\frac{G r^4}{\Big(1 + C m^2 (C c_2 + c_1 r)\Big) \left\{B + 2 \log(r) +
   2 C m^2 \Big(c_1 r + C c_2 \log(r)\Big)\right\}}} dr,\\
   &=&\sqrt{8\pi}\alpha \left(\frac{K_B}{\hbar}\right)\int_b^{b+\epsilon}\mathcal{I}(r)~dr~ \text{(say)}.
  \end{eqnarray}
\end{widetext}

  \begin{figure}[htbp]
        \includegraphics[scale=.6]{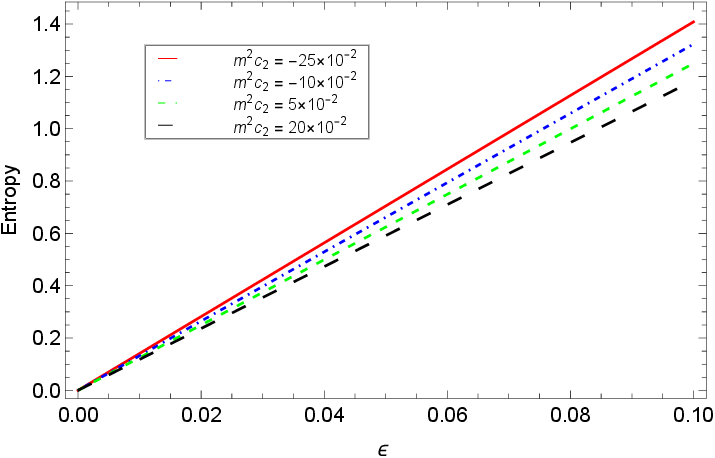}
         \includegraphics[scale=.6]{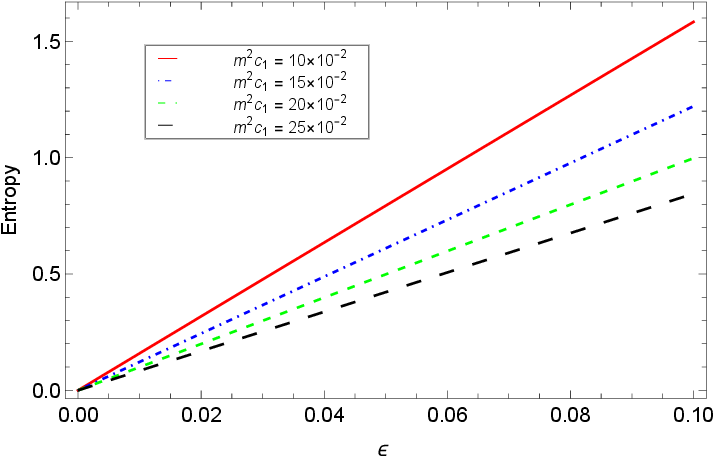}
       \caption{ Variation of the gravastar's internal entropy in relation to shell thickness.
\label{entropy}}
\end{figure}

To find the expression of the entropy explicitly, we have to perform the integral given in Eqn.~(\ref{l6}), but it is highly challenging to execute the integral given in eqn.(\ref{l6}) due to the intricacy of the formulation of $\mathcal{I}$.\\
let us assume that $h(r)$ be a function so that $\frac{dh(r)}{dr}=\mathcal{I}(r)$. After mathematical calculations, the expression of $\mathcal{I}$ becomes,
\begin{equation}\label{l1}
\mathcal{I}=h(b+\epsilon)-h(b),
\end{equation}
Taking into account the Taylor series expansion of $h(b+\epsilon)$ about `b' and taking up to second order of $\epsilon$, we finally obtain the expression of entropy as,
\begin{widetext}
\begin{eqnarray}
\mathcal{S}=\sqrt{8\pi}\alpha \left(\frac{K_B}{\hbar}\right)\left[\epsilon b^4\sqrt{\frac{G}{\big\{1 + C (b c_1 + C c_2) m^2\big\} \big[B + 2 \log(b) + 2 C m^2 \big\{b c_1 + C c_2 \log(b)\big\}\big]}}+\frac{\epsilon^2}{2}\mathcal{J}(b)\right],
\end{eqnarray}
\end{widetext}
where the expression of $\mathcal{J}(b)$ is given by,
\begin{widetext}
\begin{eqnarray}
\mathcal{J}(b)&=&\frac{1}{2 b^3 G}\Big[\frac{b^4 G}{\big\{1 + C (b c1 + C c_2) m^2\big\} \big\{B + 2 b C c_1 m^2 +
   2 (1 + C^2 c_2 m^2) \log(b)\big\}}\Big]^{\frac{3}{2}}\times\bigg[12 b^2 C^2 c_1^2 m^4 \nonumber\\&&+ 2 \big(4 B - C^2 c_2 m^2-1\big) (1 + C^2 c_2 m^2) +
  b C c_1 m^2 (12 + 7 B + 12 C^2 c_2 m^2) \nonumber\\&&+
  2 (1 + C^2 c_2 m^2) \big\{8 + C (7 b c_1 + 8 C c_2) m^2\big\} \log(b)\bigg].
\end{eqnarray}
\end{widetext}
We have shown the behavior of the entropy of the gravastar with respect to its thickness $\epsilon$ by keeping a fixed value for $m^2c_2$ in Fig.~\ref{entropy} for different values of $m^2c_1$ and vice versa. It is observed that as massive gravity parameters $m^2c_1$ and $m^2c_2$ grow, the entropy goes down but increases as $\epsilon$ increases.

\subsection{The EoS parameter}
At $r=b$, the equation of state parameter can be calculated as,
\begin{eqnarray}
\omega(b)=\frac{\mathcal{P}(b)}{\sigma(b)},
\end{eqnarray}
Now using the eqns. (\ref{sigma1}) and (\ref{p1}), we obtain the expression for $\omega(a)$ as,
\begin{eqnarray}\label{omega1}
\omega(a)=\frac{1}{2}\left[\frac{\frac{1-\frac{\mathcal{M}}{b}}{\sqrt{1-\frac{2\mathcal{M}}{b}}}-\frac{12 + 3 C m^2 (c_1 + 4 C c_2 + 2 c_1 b) - 4 k b (1 + b) \rho_c}{12\sqrt{1+m^2C(c_2 C+\frac{c_1 b}{2})-k\rho_c\frac{b^2}{3}}}}{\sqrt{1+m^2C(c_2 C+\frac{c_1 b}{2})-k\rho_c\frac{b^2}{3}}-\sqrt{1-\frac{2\mathcal{M}}{b}}}\right].\nonumber\\
\end{eqnarray}

\subsection{Energy within the thin shell}
\begin{figure}[htbp]
         \includegraphics[scale=0.6]{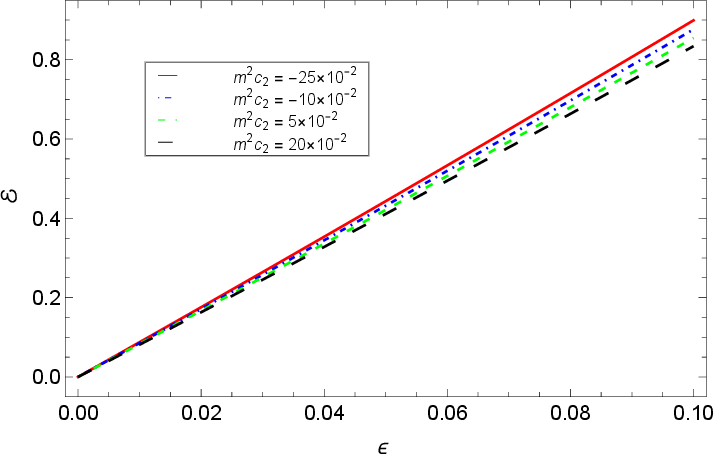}
          \includegraphics[scale=0.6]{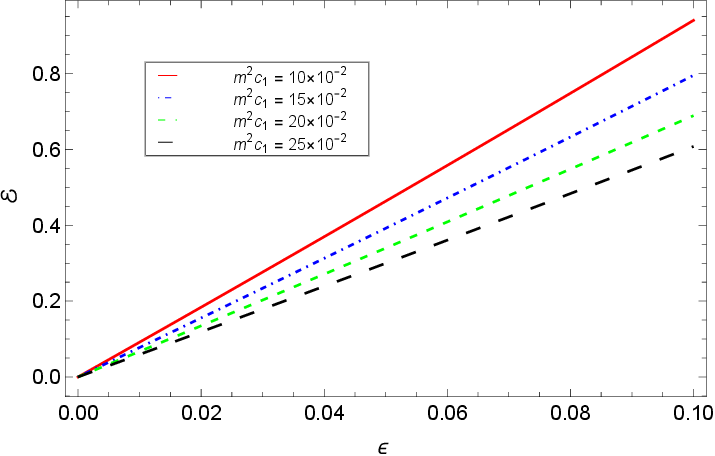}
       \caption{Variation of the gravastar's internal energy versus shell thickness. \label{energy}}
\end{figure}

The energy within the thin shell can be calculated as,
\begin{widetext}
\begin{eqnarray}\label{l3}
\mathcal{E}&=&\int_{b}^{b+\epsilon}4\pi r^2 \rho dr,\nonumber\\
&=&\frac{G \pi}{15 C^7 c_1^7 m^{14}} \Big[C c_1 m^2 r \Big\{-60 (1 + C^2 c_2 m^2)^5 +
      30 C c_1 m^2 (1 + C^2 c2 m^2)^4 r -
      20 C^2 c_1^2 m^4 (1 + C^2 c2 m^2)^3 r^2 \nonumber\\&&+
      15 C^3 c_1^3 m^6 (1 + C^2 c2 m^2)^2 r^3 -
      12 C^4 c_1^4 m^8 (1 + C^2 c_2 m^2) r^4 + 10 C^5 c_1^5 m^{10} r^5\Big\} \nonumber\\&&+
   60 (1 + C^2 c_2 m^2)^6 \log\big(1 + C m^2 (C c_2 + c_1 r)\big)\Big]_b^{b+\epsilon}
\end{eqnarray}
\end{widetext}
The graphical representation of the energy within the thin shell is shown in Fig.~\ref{energy}. Our findings in the aforementioned figures demonstrate that the energy within the shell and shell thickness have a linear connection, and that the energy tends to decrease as the associated $m^2c_2$ and $m^2c_1$ values increase, respectively in both the figures.

\section{Stability of the gravastar}\label{sec6}
\begin{figure}[htbp]
        \includegraphics[scale=.6]{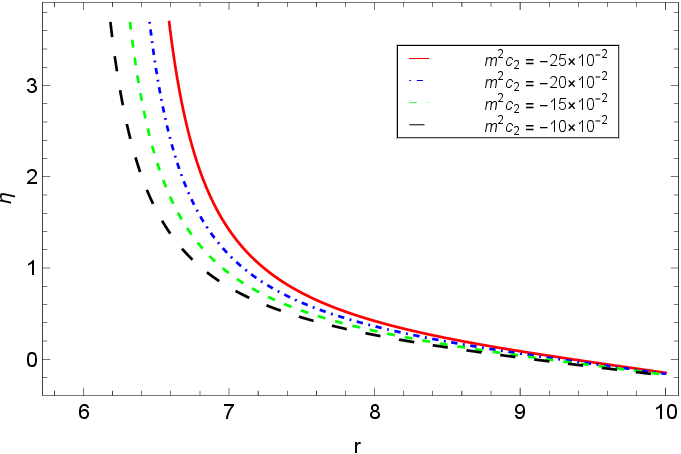}
        \includegraphics[scale=.6]{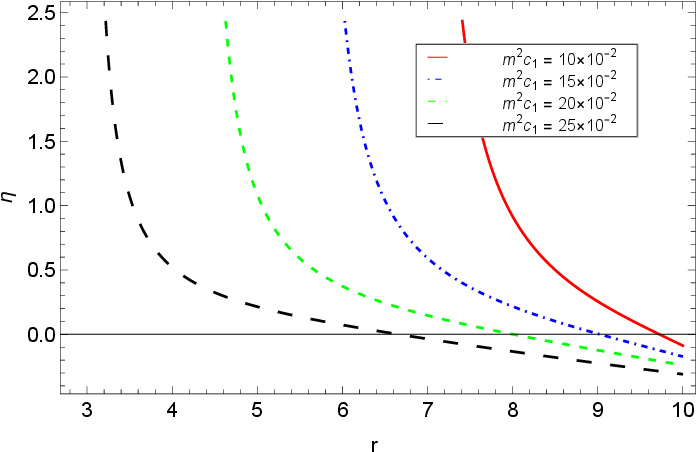}
       \caption{Stability regions of the charged gravastar. \label{eta1}}
\end{figure}

The stability of the gravastar is investigated here by using a parameter $\eta$ that can be described as the ratio of the surface pressure and surface density differentials, respectively, determined at the hypersurface $r = b$. Symbolically,
\begin{eqnarray}
\eta(b)=\frac{\mathcal{P'}(b)}{\sigma'(b)}.
\end{eqnarray}
Analysis of the stability of the thin shell wormhole under small radial perturbations has been reported by Poisson and Visser \cite{Poisson:1995sv}. They have also obtained the stability region. In the implementation of noncommutative geometry, \"Ovg\"un et al. \cite{Ovgun:2017jzt} employed the aforementioned parameter, which is essential in identifying the stability regions of the corresponding solutions of the charged thin-shell gravastar model.  By incorporating eta, Debnath also examined  the stability of charged gravastars under Rastall-Rainbow gravity in ref. \cite{Debnath:2019eor}. Using the same method, Yousaf et al. \cite{Yousaf:2019zcb} also investigated the stability of the charged gravastar in $f(R,\,T)$ gravity. This criterion was also applied by Bhar and Rej \cite{Bhar:2021oag} to investigate the stability of a charged gravastar in the presence of conformal motion within the context of $f(R,\,T)$ gravity. $\eta$ may be supposed to be understood as the speed of sound in normal scenario. In addition, normally, eta would have to fall within the interval $\eta~\in (0,\,1)$. However, the range of eta might be outside the previously stated range on the surface layer, as suggested by Poisson \& Visser \cite{Poisson:1995sv} and Lobo \& Crawford \cite{Lobo:2003xd}. For our present model, the profile of $\eta$ is shown in Fig.~\ref{eta1}. The stability region lies on the upper side of the curve as proposed by \cite{Yousaf:2019zcb,Ovgun:2017jzt}.

\begin{figure}[htbp]
        \includegraphics[scale=.7]{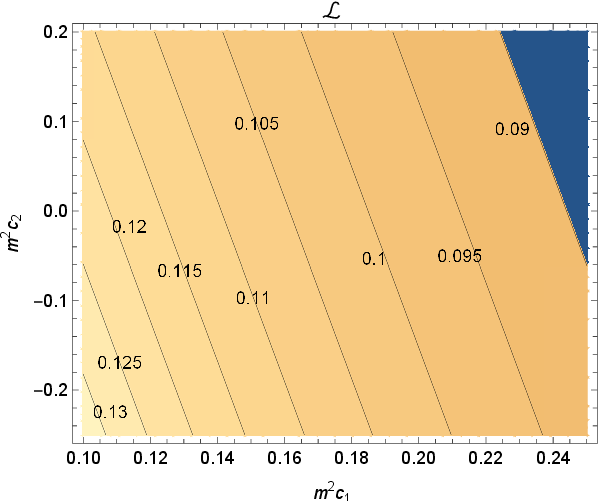}
        \includegraphics[scale=.7]{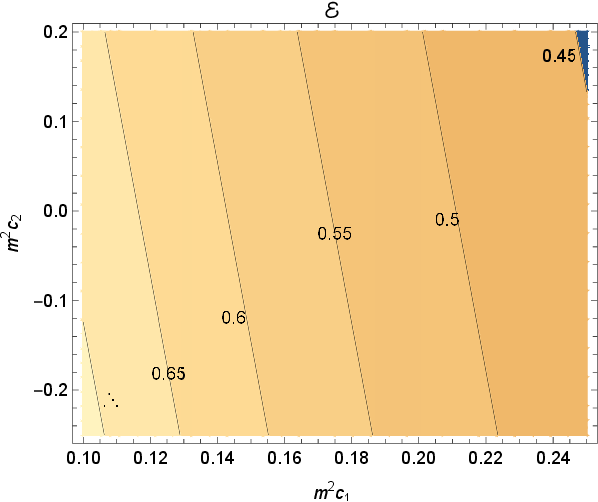}
        \includegraphics[scale=.7]{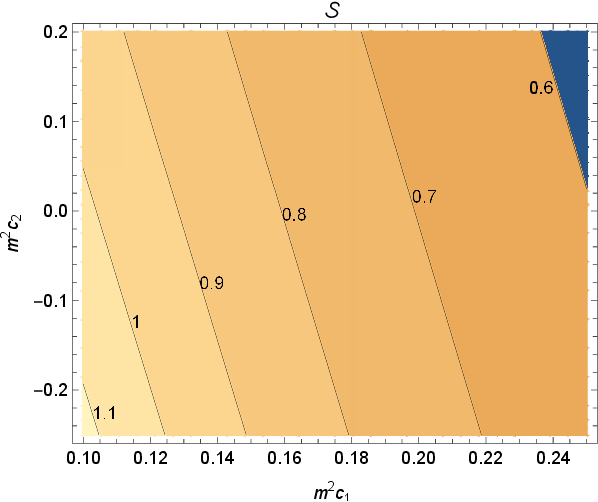}
       \caption{The contour plots of different physical properties. \label{eta}}
\end{figure}

\section{Some physical properties of Gravastar through contour plots }\label{sec7}
In this section we are interested to study some physical properties namely proper length ($\mathcal{L}$), energy content ($\mathcal{E}$) and entropy ($\mathcal{S}$) of the gravastar model with the help of contour plots in Fig.~\ref{eta} in $\left\{m^2c_1,\,m^2c_2\right\}$ plane. For all the three physical parameters, one can check that for a fixed value of $m^2c_2$ all of the physical quantities proper length ($\mathcal{L}$), energy content ($\mathcal{E}$) and entropy ($\mathcal{S}$) decrease with the increasing value of $m^2c_1$ and the same situation arises when we vary $m^2c_1$ by keeping $m^2c_2$ as fixed.

\section{Discussions and final remarks}\label{sec8}
In this work, we looked into the fundamental physical characteristics of an uncharged gravastar in the circumstances of dRGT like massive gravity. The field equations of dRGT like massive gravity are solved in three distinct regions of the gravastar by using three separate equations of state.  Our primary objective is to examine how different parameters ($C,\,m^2c_1,\,m^2c_2$) affect the gravastar structure. It has been demonstrated both analytically and graphically that the presence of massive graviton modifies the physical characteristics of the gravastar. Our study shows that in the interior region of the gravastar, the metric coefficients depend on both $m^2c_1$ and $m^2c_2$, but the massive gravity parameters have no effect on pressure and density. Conversely, in the thin shell region, the massive gravity factors influence the density, pressure, and metric coefficients. Interestingly, every solution is positive inside the gravastar's interior region and regular at the centre $r = 0$. The mathematical results demonstrate the maximum mass decrease when both $m^2c_1$ and $m^2c_2$ values increase. The highest mass might exceed $2M_{\odot}$. From the tables \ref{table1}-\ref{table2} one can note that the maximal masses are respectively $2.13~M_{\odot}$ and $2.346~M_{\odot}$ when $C =0.5,\,m^2c_1=0.1,\,m^2c_2=-0.25$, and $C =0.5,\,m^2c_2=-0.5,\,m^2c_1=0.1$.\\
We would like to draw attention to Barzegar et al.'s work in this regard. The authors of this research examine the AdS scenario for 3D gravastars within the framework of dRGT-like massive gravity. Our obtained result closely resembles that of Barzegar et al. In the current work, we have taken the thin shell approximation up to the second order, which yields findings for the physical properties of the shell that are more accurate. Interestingly, the forms of various physical quantities, such as energy, entropy, proper length, etc., have changed under the framework of dRGT like massive gravity, compared to Einstein's General Relativity. In particular, the gravastar structure reverts to the general relativity scenario when the graviton mass is zero and we restrict ourselves to the first-order approximation of the thickness parameter $\epsilon$ of the shell. From all of the aforementioned investigations, it is clear that the gravastar may exist in the background of dRGT like massive gravity.

\section*{Acknowledgments}
 P.B. is thankful to the Inter University Centre for Astronomy and Astrophysics (IUCAA), Government of India, for providing visiting associateship.

\end{document}